\documentclass[reprint,
 bibnotes,
 amsmath,amssymb,
 aps,
prb,
]{revtex4-2}

\usepackage{soul}
\usepackage{graphicx}
\usepackage{dcolumn}
\usepackage{bm}
\usepackage{xcolor}
\usepackage{amssymb}

\usepackage{subfigure}

\usepackage{mathptmx}

\usepackage{amsmath}
\DeclareMathOperator{\Tr}{Tr}

\begin{document}


\title{Conductivity and size quantization effects in semiconductor $\delta$-layer systems}

\author{Juan P. Mendez}
\email{jpmende@sandia.gov}

\author{Denis Mamaluy}
\email{mamaluy@sandia.gov}

\affiliation{Sandia National Laboratories, Albuquerque, New Mexico, 87123}%

\date{\today}

\begin{abstract}
We present an open-system quantum-mechanical 3D real-space study of the conduction band structure and conductive properties of two semiconductor systems, interesting for their beyond-Moore and quantum computing applications: phosphorus $\delta$-layers in silicon and the corresponding $\delta$-layer tunnel junctions. In order to evaluate size quantization effects on the conductivity, we consider two principal cases: nanoscale finite-width structures, used in transistors, and infinitely-wide structures, electrical properties of which are typically known experimentally. For devices widths $W<10$~nm, quantization effects are strong and it is shown that the number of propagating modes determines not only the conductivity, but the distinctive spatial distribution of the current-carrying electron states. For $W>10$~nm, the quantization effects practically vanish and the conductivity tends to the infinitely-wide device values. For tunnel junctions, two distinct conductivity regimes are predicted due to the strong conduction band quantization.

\end{abstract}

\maketitle

\section{Introduction}\label{sec:introduction}

\begin{figure}
  \centering
  \includegraphics[width=\linewidth]{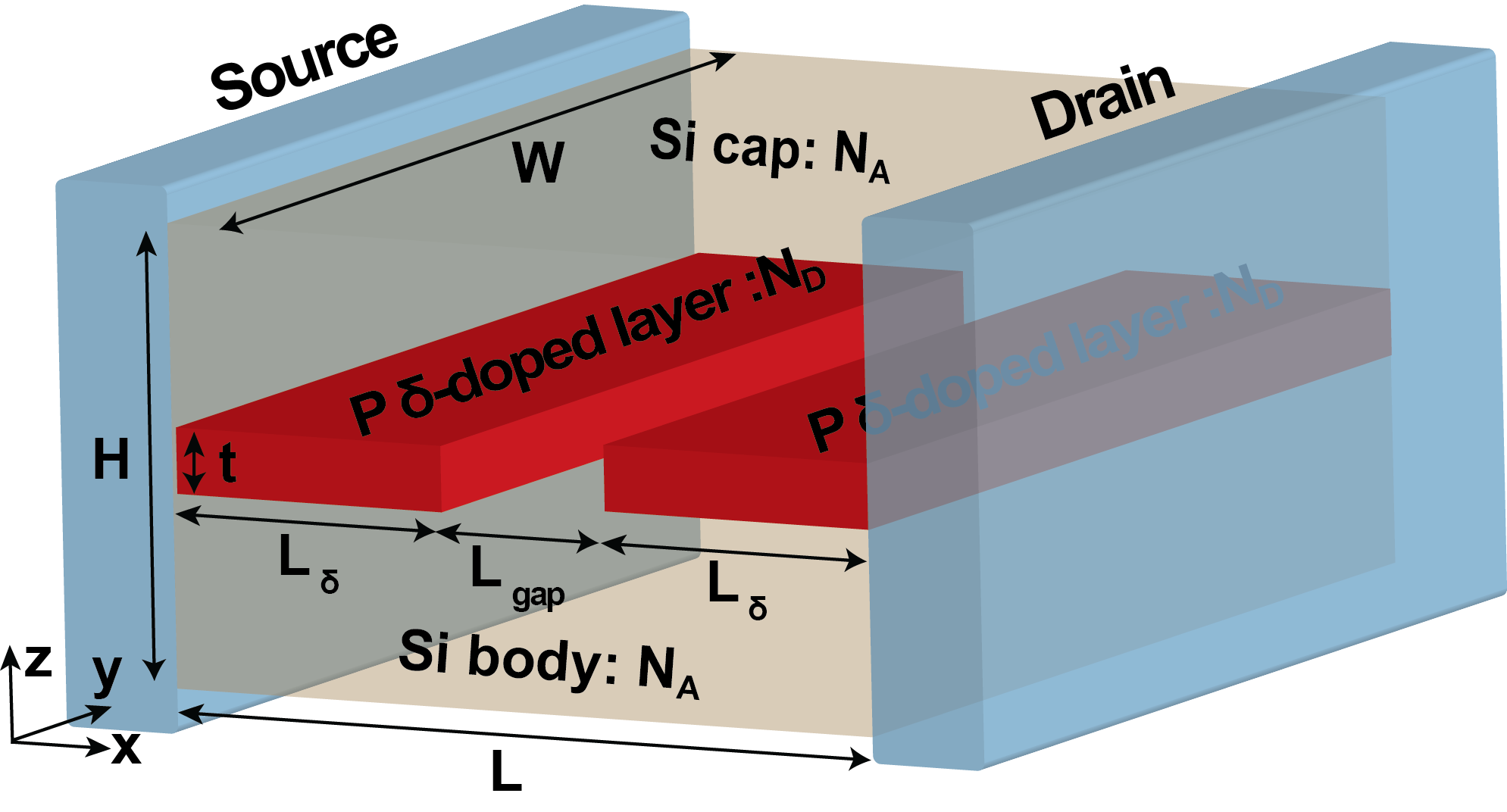}
  \caption{
  \textbf{Geometry of the Si:P $\delta$-layer tunnel junction.} Our device consists of a semi-infinite source and drain, in contact with the engineering channel. The channel is composed of a lightly doped Si body and Si cap and a very thin, highly P doped-layer with an intrinsic gap of length $L_{gap}$.
  }
  \label{fig:ideal_TJ_model}
\end{figure}

Highly conductive $\delta$-layer systems, i.e. thin, high-density layers of dopants in semiconductors are actively used as a platform  for  exploration  of  the  future  quantum  and  classical  computing  when  patterned  in  plane  with  atomic precision \cite{Skeren:2020,Bussmann:2021,Ward:2020}. Such structures, with the dopant densities above to the solid solubility limit \cite{Goh:2004}, have been shown to possess very high current densities \cite{Goh:2006,Ward:2020} and thus have a strong potential for quantum computing applications\cite{He:2019, Yang:2020} and advanced microelectronic devices \cite{Fuechsle:2012,Skere:2018}. However, at the scale important for these applications \cite{Spectrum:2021}, i.e. devices with sub-20~nm physical gate/channel lengths and/or sub-20~nm widths, that could compete with the future CMOS, the conductive properties of such systems are expected to exhibit a strong influence of size quantization effects. At the same time, experimental assessment of the conductivity of $\delta$-layer systems is typically performed using Hall measurements on samples of macroscopic dimensions ($>1$~$\mu$m) \cite{Goh:2006,Goh:2009,Reusch:2008,McKibbin:2014}. 

The electronic structure and conductive properties of Si:P $\delta$-layer systems have been a subject of previous studies based on either effective mass \cite{Drumm:2012,Mendez:2020,Mamaluy:2021,Mendez:2021}, tight-binding \cite{Lee:2011,Ryu:2013,Smith:2014,Smith:2015,Dusko:2019}, density functional theory \cite{Carter:2009,Carter:2011,Drumm:2013} formalisms or semiclassical Boltzmann theory \cite{Hwang:2013}. Recently it has been demonstrated in \cite{Mendez:2020,Mamaluy:2021,Mendez:2021} that to accurately extract the conductive properties of highly-conductive, highly-confined  systems, an open-system quantum-mechanical analysis is necessary. Such open-system treatment, that can be conducted for instance using the Non-Equilibrium Green's Function (NEGF) formalism \cite{Keldysh:1965,Datta:1997}, allows to compute the current and conductivity directly from the quantum-mechanical flux, thus avoiding semi-classical approximations, which are intrinsic to the traditional charge self-consistent closed-system or periodic boundary conditions band-structure calculation methods. It has been also demonstrated in \cite{Mamaluy:2021} that an open-system treatment in highly P doped $\delta$-layer in Si: i) permits to reveal the quantization in space and energy of the free electrons around the $\delta$-layer; ii) permits to explain the existence of the shallow 3$\Gamma$ sub-band, which has been observed experimentally \cite{Holt:2019}, without any fitting parameters; iii) predicts significant quantum-mechanical dependence of the current on the $\delta$-layer sheet thickness for a fixed dopant sheet density; and iv) provides a very good agreement with the experimental electrical measurements \cite{Goh:2006,Goh:2009,Reusch:2008,McKibbin:2014}.
 
An accurate computational description of electron tunneling in semiconductor $\delta$-layer tunnel junctions  (such the one shown in~Fig.~\ref{fig:ideal_TJ_model}) is additionally required because the tunneling rate at a $\delta$-layer junction is affected not only by the gap length and the conductivity of the $\delta$-layers, but also by quantization of the conduction electrons in energy and space \cite{Mamaluy:2021}. 
In this work we employ an efficient computational open-system quantum-mechanical treatment to explore the conductive band structure and the conductive properties of phosphorus $\delta$-layer systems in silicon (Si:P $\delta$-layer) for device widths, from nano-scale ($<20$~nm) to macro-scale ($>1$~$\mu$m) dimensions, and to analyze the influence of size quantization effects on the conductive properties for sub-12~nm device widths. All simulations are carried at the cryogenic temperature of $4$K, in which we can neglect inelastic scattering events \cite{Goh:2006,Mazzola:2014}. The analysis of the influence of different kinds of non-idealities on the tunneling current is presented in \cite{Mendez_CP:2021}. The open-system treatment is based on an application of Keldysh formalism \cite{Keldysh:1965}, known as NEGF \cite{Datta:1997}, and the effective mass theory. Generally for Si systems, the use of the effective mass approximation has been shown to be in a good agreement with tight-binding models for nanowire diameters down to 3~nm \cite{Wang:2005,Neophytou:2008}; thus, the fidelity of this approximation start to decline for very narrow systems ($<3$~nm). Here we have employed an efficient implementation of NEGF, refereed to as the Contact Block Reduction (CBR) method \cite{Mamaluy:2003,Mamaluy_2004,Mamaluy:2005,Khan:2007,Gao:2014}, which allows accurate computation of all quantum-mechanical quantities of interest (local density of states, transmission probability, current) and scales linearly $O(N)$ with the system size. We find that at the scale most interesting for applications, i.e. for device widths $W<10$~nm, quantization effects strongly affect both the conductivity and the spatial distribution of the current-carrying electron states, which, similarly to the charge distribution in the hydrogen atom, is determined by a "quantum number", i.e. the number of propagating modes. This strong spacial quantization of the current-carrying states can be utilized in novel electronic $\delta$-layer devices. Conversely, for $W>10$~nm, the quantization effects practically vanish and the conductivity tend to the values of infinitely-wide devices. Finally, two distinct conductivity regimes are predicted due to the strong conduction band quantization for tunnel junctions.

\section{Results and discussion}\label{sec:disccusion}

\subsection{Effects of the device width on the conductive properties}\label{sec:Effects of the device width $W$ on the conductive properties}

\begin{figure}
  \centering
  \includegraphics[width=\linewidth]{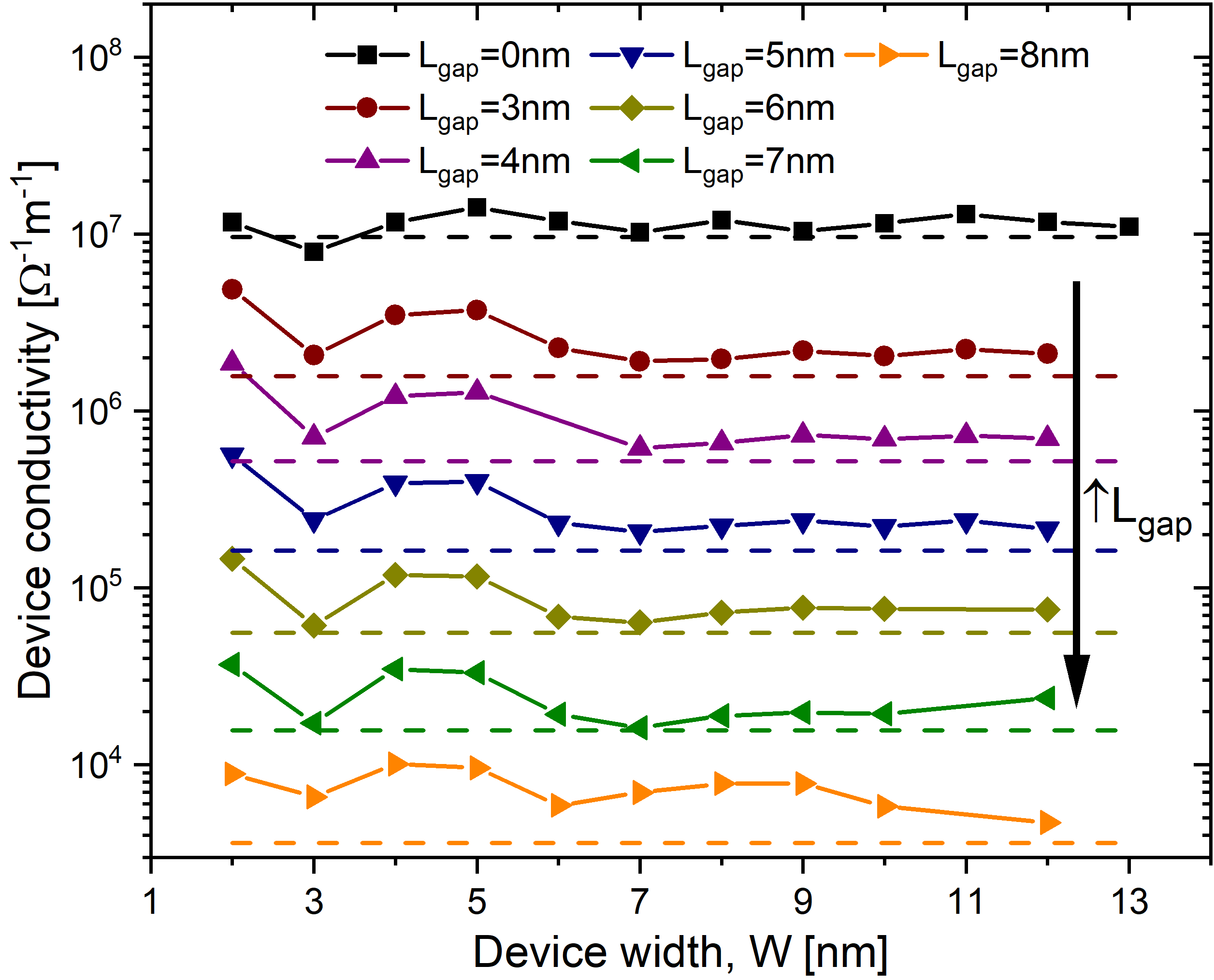} 
  \caption{
  \textbf{Conductivity for Si: P $\delta$-layer devices.} The conductivity is shown in function of the device width $W$ for different tunnel gap lengths $L_{gap}$. The corresponding conductivity values for infinitely-wide devices are indicated in dashed lines. $N_{D}=1.0 \times 10^{14}$~cm$^{-2}$, $N_{A}=5.0 \times 10^{17}$~cm$^{-3}$, $t=1.0$~nm and for an applied voltage of 1~mV.
  }
  \label{fig:conductivity vs device width}
\end{figure}

\begin{figure*}
  \centering
  \includegraphics[width=\linewidth]{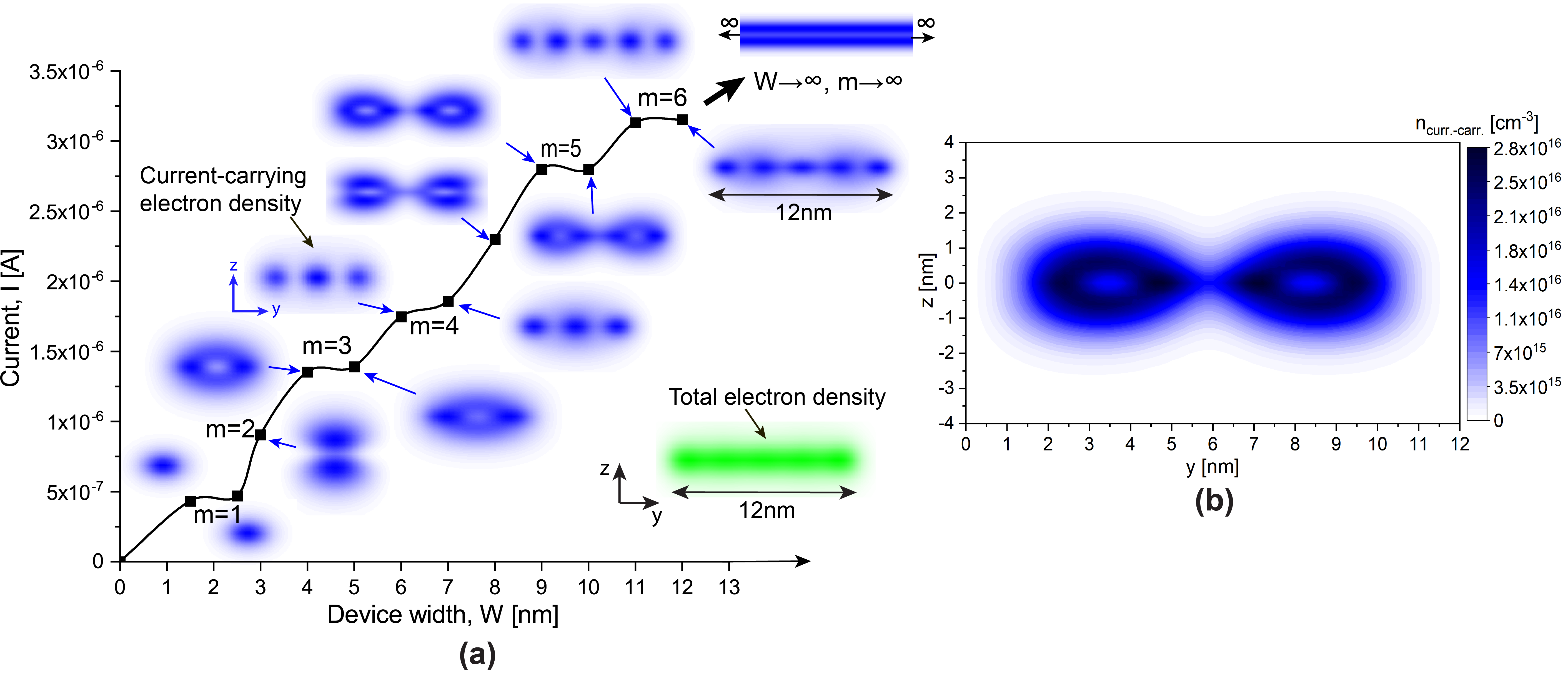} 
\caption{
\textbf{Propagation modes for Si: P $\delta$-layer systems.} \textbf{a} Current I vs device width W for $\delta$-layer systems with L$_{gap}=0$~nm: the insets in blue color show the spatial distributions of current-carrying modes across a y-z plane, indicating the corresponding number of propagating modes $m$; Inset in green color shows the total electron density that includes all (not just current-carrying) occupied electron states for a device width of $W=12$~nm (Note that $n_{total}\sim 10^{20}~\text{cm}^{-3} \gg n_{curr.-carr.}\sim 10^{16}~\text{cm}^{-3}$). \textbf{b} Detail of the spatial distribution of current-carrying states $n_{curr.-carr.}(y,z)$ for a device width  of $W=9$~nm . For all calculations, $N_{D}=1.0 \times 10^{14}$~cm$^{-2}$, $N_{A}=5.0 \times 10^{17}$~cm$^{-3}$, $t=0.2$~nm and an applied voltage of 1~mV.
}
\label{fig:current-carrying-modes}
\end{figure*}

The conductivity of infinitely-wide Si:P $\delta$-layers has been first studied  in \cite{Mendez:2020,Mamaluy:2021,Mendez:2021} using the open-system quantum-mechanical approach for infinite-width systems (see Sect. \ref{sec:methodology:from finite-width to infinite-width systems}). Here we apply the open-system treatment to more realistic/utilitarian devices of finite width and/or with an intrinsic or lightly-doped tunnel gap as shown in Fig.~\ref{fig:ideal_TJ_model}. The device consists of a semi-infinite source and drain (represented by the NEGF open boundary conditions), in contact with a channel of length $L$, which is composed of a lightly doped Si body and Si cap and a very thin, highly P doped-layer with an intrinsic gap of length $L_{gap}$, as shown in Fig.~\ref{fig:ideal_TJ_model}. In this work we consider tunnel junction devices with an intrinsic gap of length $L_{gap}$, ranging from 0~nm to 12~nm, and width $W$, ranging from 2~nm to infinity, paying a particular attention to nano-scale dimensions ($W, L_{gap}<12$~nm) which are most interesting for quantum and beyond-CMOS computing applications. The length of the channel is assumed to be $L=30~\text{nm}+L_{gap}$ to avoid the boundary effects, between the source and drain contacts, and the height of the device is chosen to be $H=12$~nm. We also assume $\delta$-layer thicknesses raging from monoatomic layer, $t=0.2$~nm, to few atomic layers, $t=1.0$~nm, a doping density of the $\delta$-layer of $N_D=1.0 \times 10^{14}$~cm$^{-2}$ and an acceptor doping densities in the Si cap and Si body of $N_A=5.0 \times 10^{17}$~cm$^{-3}$. Note that the doping density in the $\delta$-layer is given in cm$^{-2}$ (i.e. sheet doping density) to be consistent with experiments' nomenclature: $N_D^{(2D)} = t \times N_D^{(3D)}$, where $t$ is the $\delta$-layer thickness, $N_D^{(3D)}$ is the doping density in cm$^{-3}$ and $N_D^{(2D)}$ in cm$^{-2}$. 

Fig.~\ref{fig:conductivity vs device width} shows the computed conductivity in function of the device width $W$ and for different gap lengths $L_{gap}$. The dashed lines represent the conductivity values for infinitely wide devices, $W\to\infty$. Details of the computational treatment for infinitely-wide systems are presented in Sect. \ref{sec:methodology:from finite-width to infinite-width systems}. The simulations suggest that quantization due to the finite size of the device width starts to appear for widths below to 7-10~nm. This size quantization is reflected as non-monotonic increase of the conductivity. Interestingly, there is a peak in the conductivity at $W=5$~nm and a dip at $W=3$~nm. Conversely, as $W$ increases, the conductivity tends to the values for infinitely wide devices that therefore can be seen as lower bound limits. Additionally, the size quantization effect on the conductivity of $\delta$-layer tunnel structures is most notable for large tunnel gaps $L_{gap}>7$~nm. Indeed, in Fig.~\ref{fig:conductivity vs device width}, one can see that the deviation from the conductivity values for infinitely wide devices is more significant for large tunnel gaps. As we will discussed later in Section~\ref{sec:discussion:quantization effects of tunnel junction lengths}, this is due to the consequence of the quantization of the low-energy conduction band in $\delta$-layer systems and the wave-functions decoupling between the left and right $\delta$-layers in large tunnel gaps.

The oscillations of the conductivity for narrow device widths $W$ arise due to a small number of the propagating modes in a "waveguide", created by the finite-size width of the $\delta$-layer. The corresponding dependence of the current on the device width is shown in  Fig.~\ref{fig:current-carrying-modes}~\textbf{a} for a gapless $\delta$-layer structure $L_{gap}=0$. The existence of the conduction steps due to each new propagating mode is well known experimentally since 1980's \cite{Wees:1988}. Here we show, however, that in highly-confined, highly-conductive $\delta$-layer systems, the quantum number $m$, representing the number of propagating modes, determines not just the total current, but also the spatial distribution of the corresponding current-carrying electrons. The total number of propagating modes $m$ is determined by the number of peaks in the density of states (DOS) below the Fermi level or the number of steps in the electronic transmission function below the Fermi level. 

The spatial distribution of the current-carrying electron states, $n_{curr.-carr.}(y,z)$, can be obtained by performing the energy integration of the local density of electron states (LDOS) weighted by the corresponding current spectrum $i_e(E)$ as: $n_{curr.-carr.}(y,z)=\int LDOS(y,z,E)i_e(E)dE/\int i_e(E)dE$. The current spectrum and the local density of states can be obtained by expression in Eqs.~\ref{eq:current} and \ref{eq:LDOS}, respectively.
The spatial current-carrying electrons for the different modes is shown in Fig.\ref{fig:current-carrying-modes}~\textbf{a} as insets in blue color, as well as the corresponding number of propagating modes. Additionally, the total electron density is also included in the figure as an inset in green color, demonstrating only weak spatial quantization along the y-direction. However, the specific portion of electrons with energies close to the Fermi level, i.e. the current-carrying states, do exhibit a strong spatial quantization. Indeed, for $m=1$ the propagating mode reaches the maximum concentration at the center of the structure, the mode the corresponds to $m=2$ is "excited" into the further penetration along the confinement direction (z-axis), leaving the center relatively depopulated (in terms of the current-carrying states), the mode $m=3$ is again "pushed out" of the center along both z- and y- axis. One can further note that the modes $m=1,$ and $m = 4, 6,\text{etc.}$ tend to form a regular "phase" distribution of the current-carrying states (i.e. the states distributed closer to the center of the $\delta$-layer along z-axis), while the modes $m=2,$ and $m = 3, 5,\text{etc.}$ form "anti-phase" distributions (i.e. the states distributed further from the center of the $\delta$-layer along z-axis) that have the maximum current being carried in the different regions of space, separated by a few nanometers. When $W\to\infty$, the number of propagation modes in y-direction becomes infinite $m\to\infty$ as expected. Fig.~\ref{fig:current-carrying-modes}~\textbf{b} shows the density of current-carrying electron states for the anti-phase case of $\delta$-layer doping $N_{D}=10^{14}~\text{cm}^{-2}$, thickness $t=0.2$~nm (to approximate a monoatomic $\delta$-layer) and device width $W=9$~nm, which corresponds to $m=5$. As final remark, the number of propagating modes $m$ in $\delta$-layer structures is mainly determined by three factors: 1) the $\delta$-layer doping level $N_{D}$, 2) the the $\delta$-layer doping thickness $t$ and 3) the device width $W$. 

\subsection{Peculiarities of finite-width vs infinitely-wide devices}\label{sec:Peculiarities of finite-width vs infinitely-wide devices}

Next it is to get an insight into the transition from infinite-width to finite-width devices, before the quantum effects arise on the conductivity properties, i.e. the transition marked with a black arrow in the right upper part of Fig.~\ref{fig:current-carrying-modes}. In the following, therefore, we will compare the free electron energy distribution, the transmission function and the current spectrum for devices of 12~nm-width, in which the size quantum effects on the conductivity are minimum as was seen in Fig.~\ref{fig:conductivity vs device width}, and the corresponding infinitely wide counterparts. All results of infinitely wide devices are scaled to an effective device width of 12~nm for comparison purposes.  

\begin{figure}
  \centering
  \includegraphics[width=\linewidth]{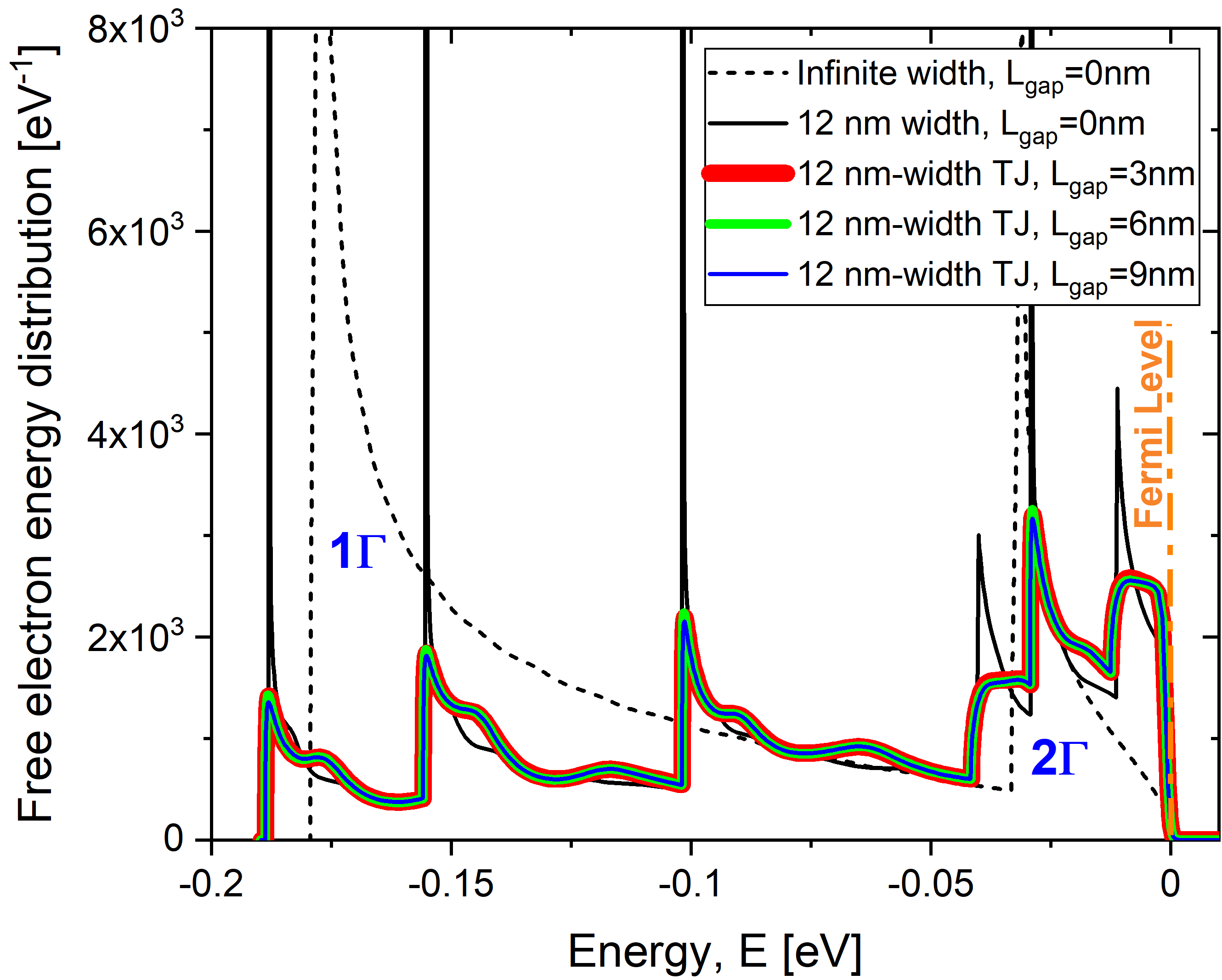}
  \caption{\textbf{Free electron energy distribution for Si:P $\delta$-layer systems.} It shows a comparison of the free electron energy distribution for infinite and 12~nm-width $\delta$-layer devices. The free electron energy distribution for the infinite-width $\delta$-layer is normalized to a width of the device of $W=12$~nm for comparison purpose. $N_{D}=1.0 \times 10^{14}$~cm$^{-2}$, $N_{A}=5.0 \times 10^{17}$~cm$^{-3}$ and $t=1.0$~nm. 
  }
  \label{fig:free_electron_distribution}
\end{figure}

Fig.~\ref{fig:free_electron_distribution} shows the free electron energy distribution of $\delta$-layer structures of 12~nm-width (in continuous lines), together with the result of infinitly wide $\delta$-layer structure with $L_{gap}=0$~nm (in dashed line), for different gap lengths $L_{gap}$. The free electron energy distribution corresponds to the total density of states multiplied by the the Fermi-Dirac distribution function, which determines the probability of a state to be occupied. First we notice that in contrast to the infinite-width $\delta$-layer systems, the finite size along the y-direction manifests itself in splitting of the occupied $1\Gamma$ sub-band, that is located around -0.18~eV, and $2\Gamma$ sub-band, that is located around -0.033~eV, into additional modes. Interestingly, for tunnel junctions ($L_{gap}>0$), two of $2\Gamma$ modes are immediately dampened as manifested in the reduction of the corresponding peaks compared to the gapless case as shown in Fig.~\ref{fig:free_electron_distribution}. In addition, as it is evident from Fig.~\ref{fig:free_electron_distribution}, the free electron distribution in tunnel junctions is almost independent of the tunnel gap length $L_{gap}$. 

\begin{figure}
  \centering
  \includegraphics[width=\linewidth]{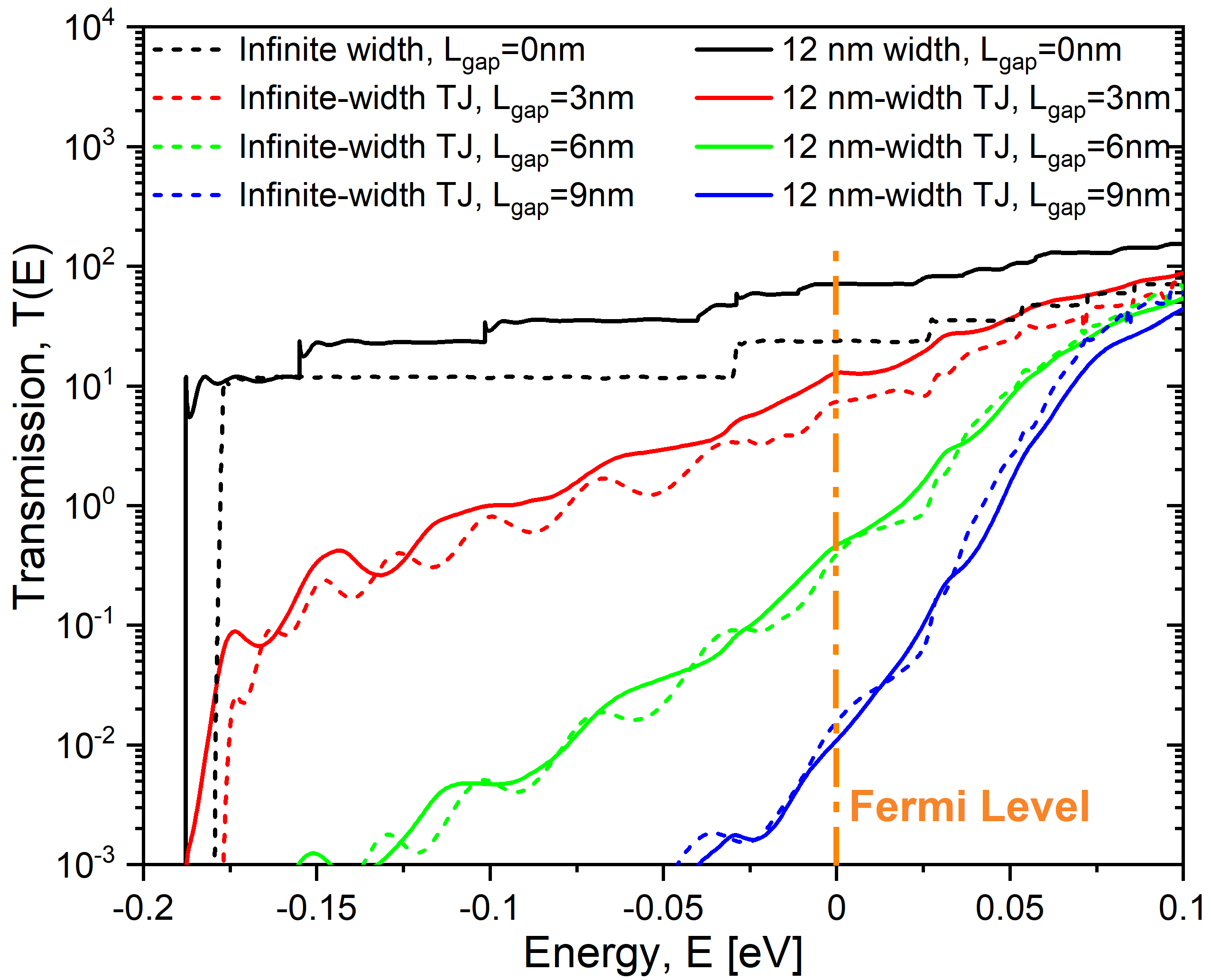}
  \caption{
  \textbf{Electronic transmission for Si:P $\delta$-layer systems.} It shows a comparison of the transmission function (logarithmic scale) for an applied voltage of $1.0$~mV for infinite and 12~nm-width $\delta$-layer devices. The transmission function for the infinite-width $\delta$-layer is normalized to a width of the device of $W=12$~nm for comparison purpose. $N_{D}=1.0 \times 10^{14}$~cm$^{-2}$, $N_{A}=5.0 \times 10^{17}$~cm$^{-3}$ and $t=1.0$~nm.
  }
  \label{fig:transmission_ideal_TJ}
\end{figure}

Fig.~\ref{fig:transmission_ideal_TJ} shows the transmission function of infinite-width devices (in dashed lines) and 12~nm-width devices (in continuous lines) for different tunnel gap lengths $L_{gap}$. The transmission function provides the sum of the probabilities for each mode of a carrier at certain given energy $E$ to carry current from the source to drain. As can be seen, the transmission function is fairly similar for both systems, infinite and finite width, especially for wider tunnel gaps. The transmission function is reduced exponentially for the low-energy modes with the length of the tunnel gap. Additionally, the transmission function strongly dependent on the gap length: the energy window in which the carrier can be transmitted is quickly reduced with the increase of the tunnel gap length.

\begin{figure}
  \centering
  \includegraphics[width=\linewidth]{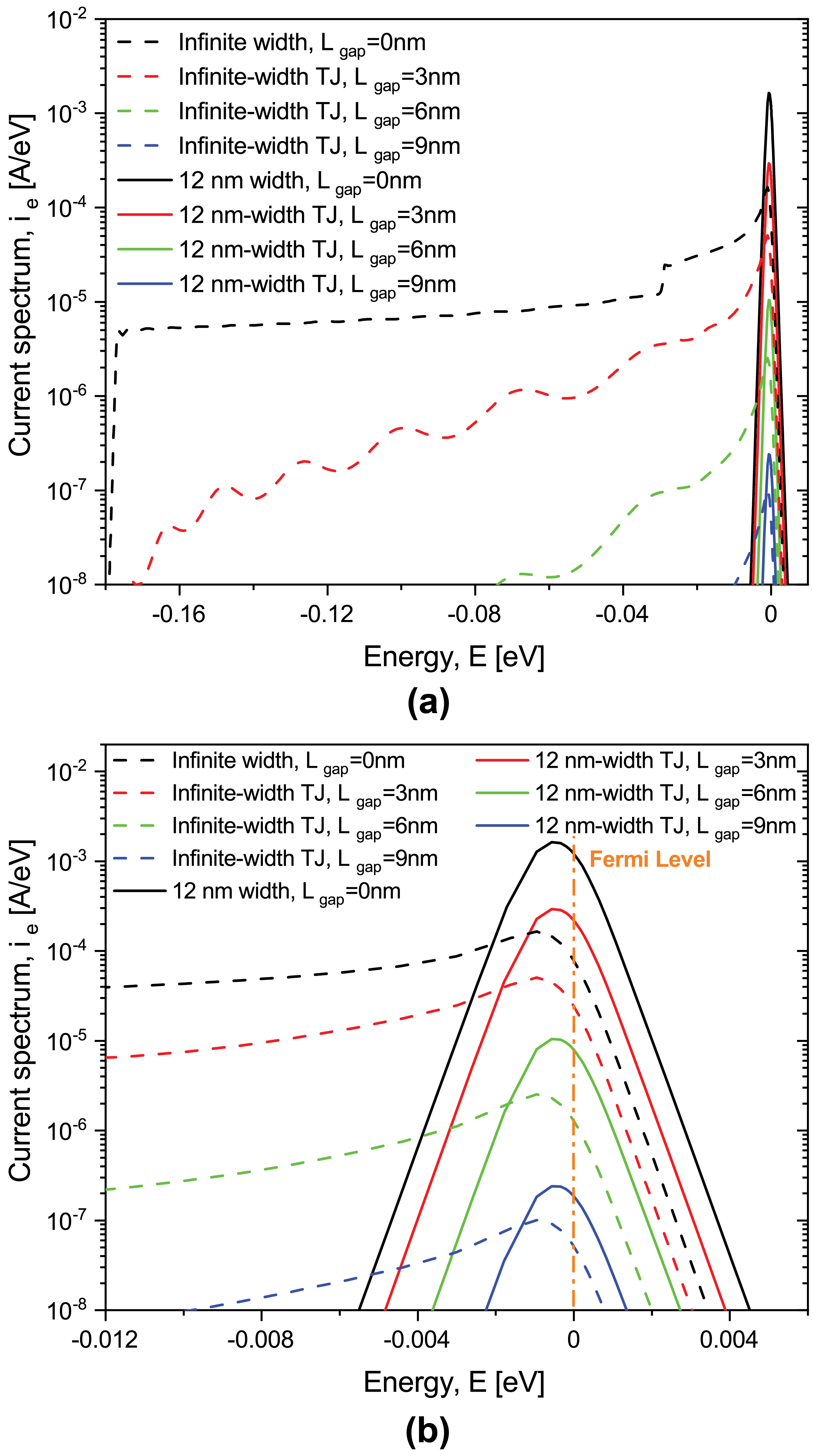}
  \caption{\textbf{Current spectrum for Si:P $\delta$-layer systems.} \textbf{a} Comparison of the current spectrum $i_e$ (logarithmic scale) for an applied voltage of $1.0$~mV between infinite and 12~nm-width $\delta$-layer devices. \textbf{b} Close-up of the current spectrum within the Fermi level. The currents for the infinite-width $\delta$-layer systems are normalized to a width of the device of $W=12$~nm for comparison purpose. $N_{D}=1.0 \times 10^{14}$~cm$^{-2}$, $N_{A}=5.0 \times 10^{17}$~cm$^{-3}$ and $t=1.0$~nm.}
  \label{fig:tunneling_current_ideal_TJ}
\end{figure}

Fig.~\ref{fig:tunneling_current_ideal_TJ} shows the current spectrum, $i_e(E)$: in dashed lines, for infinite-width devices and, in continuous lines, for 12~nm-width devices. It is evident that the current spectrum of infinite-width devices differs from the current spectrum of finite-width counterparts. The finite size along y-axis significantly affects the current spectrum limiting the current-carrying states in all cases, including the gapless case, to only a narrow energy window around the Fermi level, with the size of the energy window being proportional to the applied voltage. The mechanism of such current spectra limitation  for finite-width devices can be understood referring to the derivation in section~\ref{sec:methodology:from finite-width to infinite-width systems}: for finite-width devices, the sum in Eq.~\ref{eq:FD_sum} contains only a limited number of quantized values of $k_y$, therefore the contribution of the corresponding propagation modes with low energies $E_m$ is dampened, while only modes with the high energy (near Fermi level) can affect the current. The difference in the current spectra between finitely and infinitely-wide devices is diminished for very large tunnel junctions, as Fig.~\ref{fig:tunneling_current_ideal_TJ} indicates for $L_{gap}>9$~nm. Indeed, for sufficiently large tunnel gaps, the tunneling current spectrum is exponentially suppressed for all low-energy modes, since the effective tunnel barrier height is larger for them than for higher-energy modes. Consequently, only the modes in the immediate vicinity of the Fermi level contribute to the total current for both finite and infinite-width tunnel junctions. 

Despite the seemingly significant differences between finite-width systems and the corresponding infinitely wide counterparts, evidenced in the free electron energy distribution (see Fig.~\ref{fig:free_electron_distribution}) and in their respective current spectra (see Fig.~\ref{fig:tunneling_current_ideal_TJ}), the computed total currents, obtained by integrating the current spectra, only differ by about $20\%$. We thus conclude that while the density of states and the current spectra significantly differ, the sheet conductance (i.e. conductivity) calculations are accurate enough for finite-size systems at least down to $12$~nm-widths. At the same time, for tunnel junctions with sufficiently large gaps, both the total current and the current spectra are very similar between infinite-width and finite-width systems, since in both cases the current is carried only by the high-energy electrons in the vicinity of system's Fermi level.

\subsection{Quantization effects in $\delta$-layer tunnel junctions}\label{sec:discussion:quantization effects of tunnel junction lengths}

\begin{figure}
  \centering
  \includegraphics[width=\linewidth]{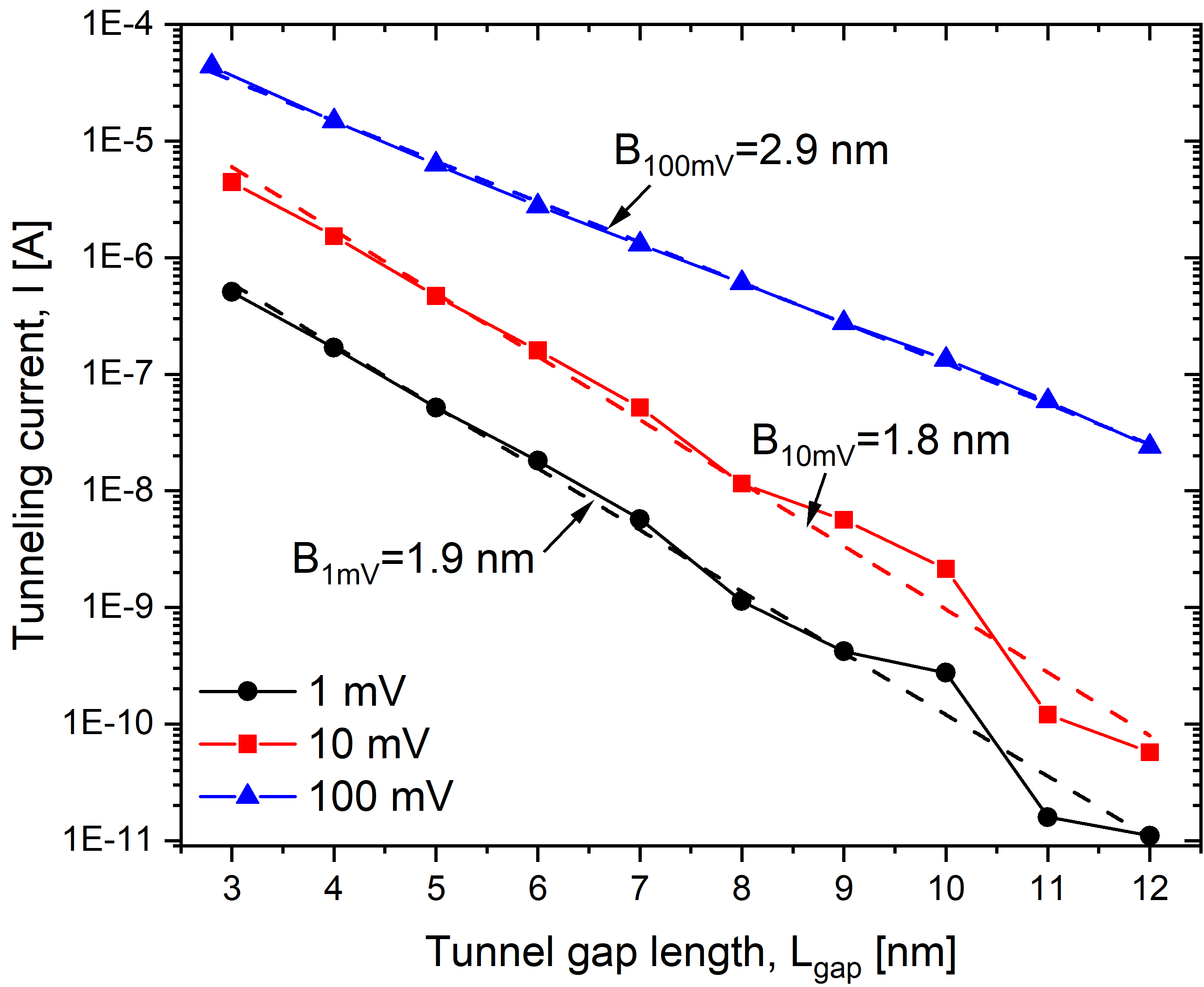}
  \caption{
  \textbf{Characteristic tunneling current curves.} The total current (logarithmic scale) $I$ vs. tunnel gap length $L_{gap}$ for three different applied voltages of $V=1~\text{mV}, 10~\text{mV}$ and $100~\text{mV}$ is shown. 
  $N_D=1.0 \times 10^{14}$~cm$^{-2}$ and $N_A=5.0 \times 10^{17}$~cm$^{-3}$, $W=12$~nm and $t=1$~nm. Dashed lines represent least-square fits to the exponential trend.
  }
  \label{fig:I_vs_W_ideal_TJ}
\end{figure}

\begin{figure*}
  \centering
  \includegraphics[width=\linewidth]{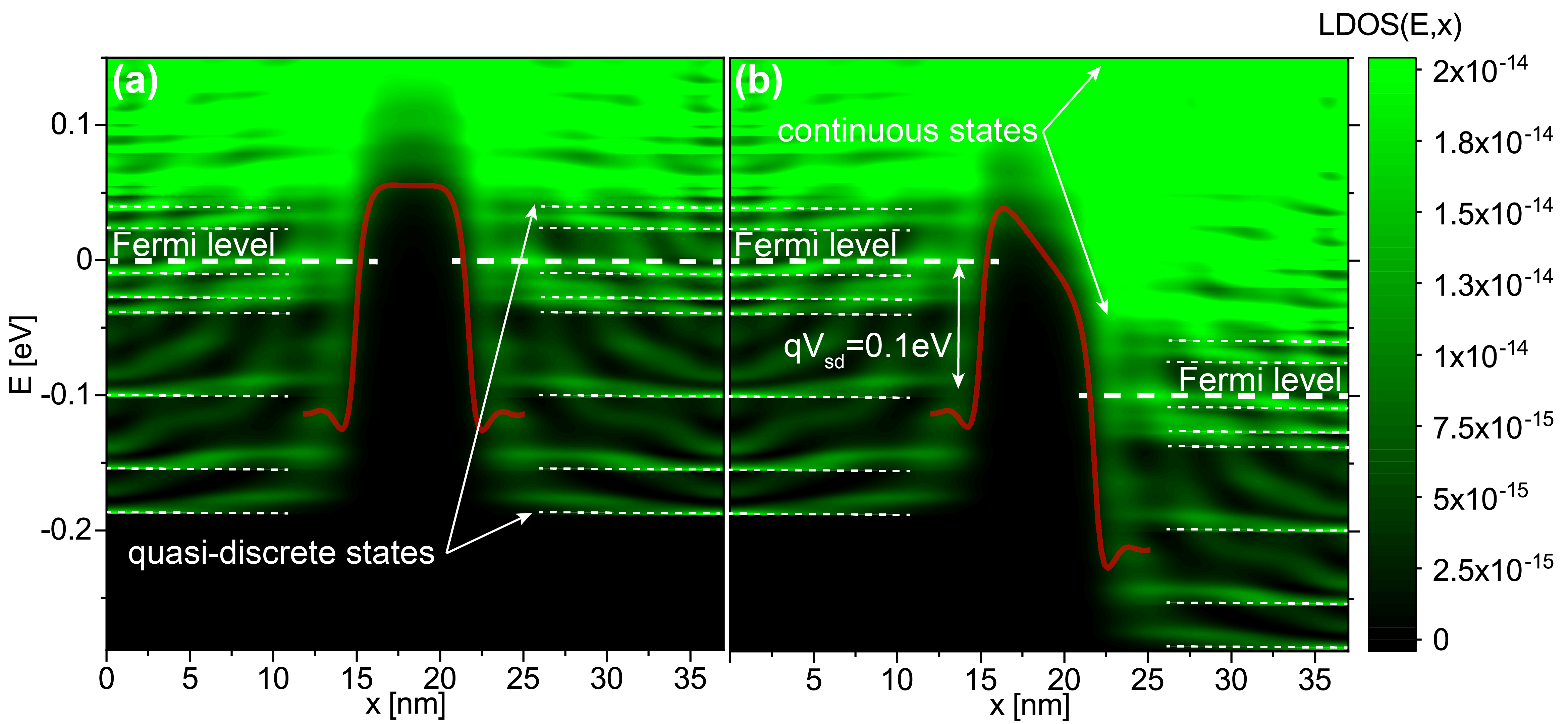} 
  \caption{
  \textbf{Local Density of States for $\delta$-layer tunnel junctions.}
  The $LDOS(E,x)$ for a tunnel junction of $L_{gap}=7$~nm is shown in \textbf{a} and \textbf{b} when a voltage of 1~mV and 100~mV is applied to the right side of the tunnel junction, respectively. Subfigure \textbf{a} indicates that the quantized states around the Fermi level might affect the conductivity for low applied biases due to a possible mismatch between the left and right quasi-discrete states, i.e., when the corresponding quasi-discrete peaks from the left and right sides are not aligned. Subfigure \textbf{b} shows that for high applied voltages the mismatch becomes impossible due to the availability of unoccupied continuum states on the right side for tunneling from the left side. In \textbf{a} and \textbf{b}, the corresponding effective 1D potentials are also shown, calculated by integrating over the (y,z)-plane the actual charge self-consistent 3D potentials weighted with the electron density.
  $N_D=1.0 \times 10^{14}$~cm$^{-2}$, $N_A=5.0 \times 10^{17}$~cm$^{-3}$, $W=12$~nm and $t=1$~nm.
  }
  \label{fig:LDOS_x}
\end{figure*}

The tunneling current vs the tunnel gap length $L_{gap}$ is shown in Fig.~\ref{fig:I_vs_W_ideal_TJ} for three different voltages $V=1$~mV, $10$~mV and $100$~mV. In the range of gap lengths $L_{gap}=0,...,7$~nm, the I vs $L_{gap}$ trend is practically exponential for all three voltages, i.e. $\ln{I}\sim\ln{I_{L_{gap}=0}}-L_{gap}/B_{voltage}$, where $I_{L_{gap}=0}$ is the current when $L_{gap}=0$, $L_{gap}$ is the tunnel gap length and $B_{voltage}$ is a proportional constant related with the barrier height. Indeed, in this range of gap values we get $B_{1-10~\text{mV}}\approx1.9~\text{nm}$ and $B_{100~\text{mV}}=2.9~\text{nm}$, which can be translated into the effective barrier heights of an equivalent rectangular barrier ($L_{gap}\times h_{voltage}$), where $h_{voltage}=d_{val}\hbar^{2}/(8B_{voltage}^{2}m_{t})$, $\hbar$ is the reduced Planck's constant, $d_{val}$ is the valley degeneracy, and $m_t$ is the transverse effective electron mass. Substituting the corresponding $B_{voltage}$ values, we obtain $h_{1-10\text{mV}} \approx 80~\text{meV}$ and $h_{100\text{mV}} \approx 30~\text{meV}$, which agrees well with the effective barrier height shown in Fig.~\ref{fig:LDOS_x} for a tunnel gap of $L_{gap}=7$~nm, i.e. the barrier height from the Fermi level to the maximum value of the effective 1D electrostatic potential, computed using our charge self-consistent scheme. However, a deviation from the exponential trend can be seen for large tunnel gaps, $L_{gap}>7$~nm, and it is the consequence of the quatization of the low-energy conduction band\cite{Mamaluy:2021} and the resulting mismatch between occupied and unoccupied quasi-discrete electron states in the left and right $\delta$-layers, respectively, at low bias. Fig.~\ref{fig:LDOS_x} shows the LDOS along x-direction for a tunnel junction of $L_{gap}=7$~nm, i.e. the available states which can be occupied by the free electrons in space-energy dimension; in the low-temperature regime, only the states below the Fermi level are occupied. As one might observe from Fig.~\ref{fig:LDOS_x}, the low-energy LDOS are quasi-quantized, highlighted with dashed lines in the figure, however, for high energies, the states are not quantized anymore, the LDOS are practically continuous in space-energy, as can be seen above the Fermi level for high energies. When the occupied quasi-discrete states from the left side overlap most with the unoccupied quasi-discrete states (corresponding to the ones above the Fermi level) from the right side, it results in a considerably increase of the tunneling current as shown in Fig.~\ref{fig:I_vs_W_ideal_TJ} for $L_{gap}=10$~nm; on the contrary, if the overlap is reduced, as it happens for $L_{gap}=11$~nm, there is a reduction of the current as the result of the mismatch. For low biases ($\lesssim 10~\text{mV}$), the mismatch can only exist for sufficiently large tunnel gaps, $L_{gap}>7$~nm, because the existing coupling of the left and right $\delta$-layer wave-functions for narrow tunnel gaps ($L_{gap}<7$~nm) effectively equalizes the electron states on both sides, thus reducing this mismatch. Conversely, the quantization effect of the conduction band on the conductivity vanishes for high applied voltages ($\gtrsim 100~\text{mV}$) as shown in Fig.~\ref{fig:I_vs_W_ideal_TJ} for an bias of $100$~mV. When a high bias is applied, e.g. to the right side (or drain) of the tunnel junction, it effectively makes the unoccupied high-energy  continuous states of the right side available for the tunneling from the left side, thus negating the effects of the quantization of the occupied left side states on the current, as can be seen in Fig.~\ref{fig:LDOS_x}~\textbf{b}.

\subsection{Conclusions}\label{sec:final remark}

In this work, we have used an efficient open-system quantum-mechanical treatment to explore the conductive band structure and the conductive properties in Si:P $\delta$-layers for device widths ranging from nano-scale to macro-scale dimensions, and analyze the influence of size quantization effects on the conductive properties for sub-12~nm device widths. For device widths $W<10$~nm, quantization effects strongly affect not only the conductivity, but also the spatial distribution of the current-carrying electron states. Conversely, for $W>10$~nm, the quantization effects practically vanish and the conductivity tends to the values of infinitely-wide devices. Additionally, we have revealed and discussed the mechanism of two conductivity regimes in $\delta$-layer tunnel junctions with $L_{gap}>7$~nm: a low-voltage regime, where conduction band quantization effects play a very significant role, and a high-voltage regime, where the quantum effects on the current are negligible.

Finally, we point out that the strong spacial quantization of the current-carrying states could be utilized in novel electronic $\delta$-layer switches, where the number of propagating modes and their match/mis-match could be controlled by external electric fields, thus strongly affecting the current. In regular $\delta$-layer conductors the particular distribution of current-carrying states directly affects their penetration depth into Si body and cap, which typically has a large concentration of impurities (see e.g. \cite{Ward:2020}). Thus, the control of the number of propagating modes may give an additional degree of control over the rate of impurity scattering. 

\section{Method}\label{sec:methodology}

\subsection{Open-system treatment}\label{sec:methodology:open-system}

The open-system framework used in this work is based on an application of Keldysh formalism \cite{Keldysh:1965}, known as Non-Equilibrium Green Function (NEGF) \cite{Datta:1997} (Sect.~\ref{sec:methodology:NEGF}), and the effective mass theory (Sect.~\ref{sec:methodology:device Effective-Mass Hamiltonian}). The NEGF formalism, described in Sect.~\ref{sec:methodology:NEGF}, defines the Green's function matrix through the inverse $G(E)=[1-H_{0}-\Sigma(E)]^{-1}$, which allows to compute the electron density and current through a charge self-consistent scheme as described in Sect.~\ref{sec:methodology:charge self-consistency}. However, the computation of an inverse matrix generally also scales as $O(N^3)$, with $N$ being the size of the device Hamiltonian matrix $H_{0}$ (Sect.~\ref{sec:methodology:device Effective-Mass Hamiltonian}). For instance, for a device of dimensions $12\text{nm}\times15\text{nm}\times50\text{nm}$ with a uniform grid of 0.2~nm, $N>10^6$ and $N^3>10^{18}$, making a direct computation of the inverse very challenging. Here we have employed an efficient implementation of NEGF, refereed to as the Contact Block Reduction (CBR) method, which allows accurate computation of all quantum-mechanical quantities of interest (the local density of states, transmission probability, current) and scales linearly $O(N)$ with the system size. The CBR method in 3D real-space is summarized in Sects.~\ref{sec:methodology:CBR} and \ref{sec:methodology:incomplete set of eigenstates}.

  \subsection{Non-Equilibrium Green's Function Formalism}\label{sec:methodology:NEGF}
  
The current for two-contact device $(D)$ from source $(s)$ to drain $(d)$ can be computed within the Landauer-Buttiker formalism through the transmission function
\begin{equation}\label{eq:current}
I_{sd} = \int i_e(E) dE = \frac{2e}{h} \int T_{sd}(E)(f_{s}(E)-f_{d}(E))dE,
\end{equation}
where $i_e(E)$ is the current spectrum, $e$ is the electron charge, $h$ is the Planck's constant, $E$ is the energy, $f_{s(d)}(E)=f(E+qV_{s(d)})$ is the Fermi-Dirac distribution function within source (drain), to which a voltage $V_{s(d)}$ is applied, and $T_{sd}(E)$ is the electronic transmission from source to drain. Within the Green's function formalism \cite{Datta:1997}, the transmission function is given by
\begin{equation}
T_{sd}(E) = \Tr(\bm{\Gamma}_{s}\bm{G}_D\bm{\Gamma}_{d}\bm{G}^{\dag}_D)
\end{equation}
with
\begin{equation}
\bm{\Gamma}_{s(d)}=i(\bm{\Sigma}_{s(d)}-\bm{\Sigma}_{s(d)}^{\dag})
\end{equation}
where $\bm{\Gamma}_{s(d)}$ are the coupling ($N_D \times N_D$)-matrices between the device and the source (drain), $\bm{\Sigma}_{s(d)}$ are the self-energy ($N_D \times N_D$)-matrices, which describe the effects of the source (drain) on the electronic structure of the device by providing the appropriate open-system boundary conditions, $\bm{G}_D$ and $\bm{G}_D^{\dag}$ are the retarded and advanced Green's function ($N_D \times N_D$)-matrices of the coupled device with the source and drain (open-system device), respectively. $N_D$ is the total number of grid points of the discretized device domain. The retarded Green's function matrix is defined by
\begin{equation}\label{eq:Grd}
\bm{G}_D
=[\bm{E}^{+}-\bm{H}_D^0-\bm{\Sigma}]^{-1}
=[ \bm{E}^{+} - \bm{H}_D^0 - \bm{\Sigma}_s - \bm{\Sigma}_d ]^{-1},
\end{equation}
where $\bm{H}_D^0+\bm{\Sigma}_s+\bm{\Sigma}_d$ is the non-Hermitian effective Hamiltonian ($N_D \times N_D$)-matrix of the open device. Eq.~\ref{eq:Grd} can be also expressed in terms of the Green's function of the decoupled device (i.e. the closed system) $\bm{G}^0=[\bm{E}^{+} - \bm{H}_{D}^0]^{-1},$ where $\bm{E}^{+} = \bm{I}(E + i\epsilon),~\text{with}~\epsilon \to 0+$, via the Dyson equation:
\begin{equation}
\bm{G}_D=[\bm{I}-\bm{G}_D^0\bm{\Sigma}]^{-1} \bm{G}_D^0.                           
\end{equation}
The decoupled device Green function $\bm{G}_D$ can be computed using its spectral representation:
\begin{equation}
\bm{G}_D^0 = [\bm{E}^{+}-\bm{H}_D^0]^{-1}=\sum_{\alpha} \frac{|\bm{\psi}_{\alpha} \rangle \langle \bm{\psi}_{\alpha} |}{E^+-E_{\alpha}},
\end{equation}
and $E_\alpha$ and $| \bm{\psi}_{\alpha} \rangle$ are the eigenvalues and eigenvectors of $\bm{H}_D^0\bm{\psi}_{\alpha}=E_{\alpha}\bm{\psi}_{\alpha}$.
The electron density $n(\bm{r}_i)$ is given by

\begin{equation}\label{eq:LDOS}
\begin{aligned}
n(\bm{r}_i)= &\int_{-\infty}^{\infty} LDOS(\bm{r}_i,E)dE \\\nonumber 
           = &\int_{-\infty}^{\infty} \frac{1}{2\pi}\langle\bm{r}_i|\bm{G}_D\bm{\Gamma}_{s}\bm{G}^{\dag}_D|\bm{r}_i\rangle f_{s}(E)dE + \\\nonumber              
             &\int_{-\infty}^{\infty} \frac{1}{2\pi}\langle\bm{r}_i|\bm{G}_D\bm{\Gamma}_{d}\bm{G}^{\dag}_D|\bm{r}_i\rangle f_{d}(E)dE.
\end{aligned}
\end{equation}
where $LDOS(\bm{r}_i,E)$ is the local density of states. Note that the dimension of $\bm{G}_D$ and $\bm{\Gamma}_{\lambda}$ is ($N_D \times N_D$).

  \subsection{Contact Block Reduction method}\label{sec:methodology:CBR}
  
\begin{figure}
  \centering
  \includegraphics[width=0.8\linewidth]{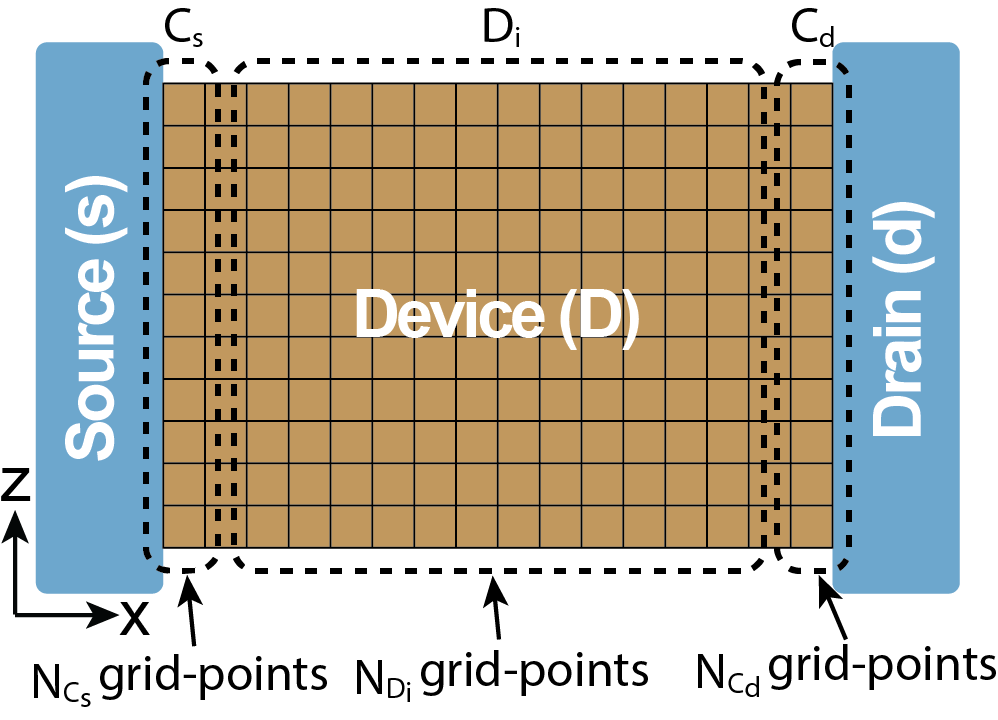}
  \caption{
  \textbf{Schematic computational model for the CBR method.} The two-contacts model is composed of two semi-infinite contacts, source (s) and drain (d), and a channel or device (D).
  }
  \label{fig:CBR_model}
\end{figure}

We next review the Contact Block Reduction (CBR) method presented in \citeauthor{Mamaluy:2003}~\cite{Mamaluy:2003,Mamaluy_2004,Mamaluy:2005}, which allows a very efficient calculation of the density matrix, transmission function, etc. of an arbitrarily shaped, multiterminal two- or three-dimensoinal open device within the NEGF formalism. In the following we apply the CBR method to the two-contact device shown in Fig.~\ref{fig:CBR_model}. The device consists of two semi-infinite contacts, source (s) and drain (d), which are in contact to the engineering channel, named as device (D).

We first discretize the domain of the device in $N_D$ grid-points, and subdivide them into $N_C=N_{C_s}+N_{C_d}$ boundary grid-points and into $N_{D_i}=N_D-N_C$ interior grid-points. $N_{C_{s(d)}}$ corresponds to the boundary grid-points between the device and source (drain). Furthermore, we assume that the real-space Hamiltonian matrix that corresponds to this discretization only couples sites within some finite range with one another, typically first nearest-neighbors. The total grid-points in the device domain is described by the following set
\begin{equation}
\begin{aligned}
\Omega &= \{ C \cup \{\textrm{interior of device} \} \} = \{ C_{s} \cup C_{d} \cup D_i \} \\
       &=\{ \{1,...,N_{C_s} \} \cup \{ N_{C_s}+1,...,N_{C_s}+N_{C_d} \} \cup \{ N_{C}+1,...,N_D \} \}
\end{aligned}
\end{equation}

With this discretization of the device domain, the self-energy matrices $\bm{\Sigma}_{s(d)}$, which represent the coupling of the source (drain) to the device, is given by
\begin{equation}
\bm{\Sigma}_{s(d)} (i,j)=
    \begin{cases}
    	[\bm{W}_{Ds(d)} \bm{G}_{s(d)}^0 \bm{W}^{\dag}_{Ds(d)} ]_{ij}, \quad &i,j \in    C_{s(d)} \\
    	0,                                                            \quad &i,j \notin C_{s(d)} \\
    \end{cases}  
\end{equation}
where $\bm{W}_{Ds(d)}$ are the Hamiltonian coupling matrices between the device $D$ and contact $s(d)$ and $\bm{G}_{s(d)}^0$ are the retarded Green's function matrix of decoupled source (drain). The total self-energy matrix ($N_D \times N_D$)-$\bm{\Sigma}$ can be expressed by the following block-diagonal matrix of the form
\begin{equation}
\bm{\Sigma} = \bm{\Sigma}_s + \bm{\Sigma}_d =
   \begin{pmatrix}
      \bm{\Sigma}_{C_s}  & \bm{0}           & \bm{0} \\
      \bm{0}           & \bm{\Sigma}_{C_d}  & \bm{0} \\
      \bm{0}           & \bm{0}             & \bm{0} \\      
   \end{pmatrix},
\end{equation}
where
\begin{equation}
\bm{\Sigma}_C = 
  \begin{pmatrix}
      \bm{\Sigma}_{C_s}  & \bm{0}           \\
      \bm{0}           & \bm{\Sigma}_{C_d}  \\
   \end{pmatrix},
\end{equation}
is of dimension of ($N_C \times N_C$).
Analogously, the retarded Green's function matrix for the device can be also rewritten by the following submatrices
\begin{equation}
\bm{G}_D =
   \begin{pmatrix}
      \bm{G}_C       & \bm{G}_{CD_i}    \\
      \bm{G}_{D_iC}  & \bm{G}_{D_i}     \\
   \end{pmatrix},
\end{equation}
where $\bm{G}_C$ is of dimension of ($N_C \times N_C$), $\bm{G}_{D_i}$ is of dimension of ($N_{D_i} \times N_{D_i}$) and $\bm{G}_{CD_i}$ is of dimension of ($N_C \times N_{D_i}$). 

Using the new representation for the above matrices, the Dyson equation can be then rewritten as
\begin{equation}
\bm{G}_D = \bm{A}^{-1} \bm{G}_D^0=[\bm{I}-\bm{G}_D^0\bm{\Sigma}]^{-1} \bm{G}_D^0, 
\end{equation}
where the matrices have the following forms:
\begin{equation}
\bm{G}_D^0 =
   \begin{pmatrix}
      \bm{G}_{C}^0    & \bm{G}_{CD_i}^0  \\
      \bm{G}_{D_iC}^0 & \bm{G}_{D_i}^0  \\
   \end{pmatrix}
\end{equation}
and
\begin{equation}
\bm{A} =
   \begin{pmatrix}
      \bm{A}_{C} & \bm{0}  \\
      \bm{R}     & \bm{1}  \\
   \end{pmatrix}.
\end{equation}
The submatrices $\bm{A}_C=\bm{1}-\bm{G}^0_C\bm{\Sigma}_C$ and $\bm{G}_{C}^0$ are of dimension of ($N_C \times N_C$), the submatrices $\bm{R}=-\bm{G}_{D_iC}^0\bm{\Sigma}_C$ and $\bm{G}_{D_i}^0$ are of dimension of ($N_{D_i} \times N_C$), and $\bm{G}_{CD_i}^0$ is of dimension of ($N_C \times N_{D_i}$).
Thus, the retarded Green's function of the open system is given by 
\begin{equation}
\begin{aligned}
\bm{G}_D &=  
   \begin{pmatrix}
      \bm{G}_{C}     &  \bm{G}_{CD_i}  \\
      \bm{G}_{D_iC}  &  \bm{G}_{D_i}   \\
   \end{pmatrix} \\\nonumber                      
&=
   \begin{pmatrix}
      \bm{A}_{C}^{-1}\bm{G}_C^0                         & \bm{A}_{C}^{-1}\bm{G}_{CD_i}^0  \\
      -\bm{R}\bm{A}_{C}^{-1}\bm{G}_C^0+\bm{G}_{D_iC}^0  & -\bm{R}\bm{A}_{C}^{-1}\bm{G}_{CD}^0+\bm{G}_{D_i}^0  \\
   \end{pmatrix}
\end{aligned}
\end{equation}
and the Green's function within the contact region by
\begin{equation}
\bm{G}_C = \bm{A}_{C}^{-1}\bm{G}_C^0 = [\bm{I}-\bm{G}_C^0\bm{\Sigma}_{C}]^{-1} \bm{G}_C^0. 
\end{equation}

The electron transmission from source $s$ to drain $d$ can be rewritten as
\begin{equation}\label{eq:transmission}
T_{sd}(E) = \Tr(\bm{\Gamma}_{C_s}\bm{G}_C\bm{\Gamma}_{C_d}\bm{G}^{\dag}_C)
\end{equation}
where
\begin{equation}\label{eq:transmission2}
\bm{\Gamma}_{C_{s(d)}}=i(\bm{\Sigma}_{C_{s(d)}}-\bm{\Sigma}_{C_{s(d)}}^{\dag}).
\end{equation}
We note that the dimension of all submatrices involved in Eqs.~\ref{eq:transmission} and \ref{eq:transmission2} is of ($ N_C \times N_C$).

Similarly, the electron density can also be written as
\begin{equation}\label{eq:density}
\begin{aligned}
n(\bm{r}_i) = \sum_{\alpha,\beta} \int_{-\infty}^{\infty} &
\bigg( \langle\bm{r}_i\big|\alpha\rangle \langle\beta\big|\bm{r}_i\rangle \Xi^{s}_{\alpha\beta}(E) f_{s}(E)  \\\nonumber
& +  \langle\bm{r}_i\big|\alpha\rangle \langle\beta\big|\bm{r}_i\rangle \Xi^{d}_{\alpha\beta}(E) f_{d}(E) \bigg) dE
\end{aligned}
\end{equation}
where the density matrix due to source(drain), $\Xi^{s(d)}_{\alpha,\beta}(E)$, is given in the closed-system basis (indexed by $\alpha,\beta$)
\begin{equation}
\Xi^{s(d)}_{\alpha,\beta}(E)=\frac{1}{2\pi}\frac{\Tr \bigg[ \big[\big|\beta\rangle \langle\alpha\big| \big]_{C} \bm{B}_C^{-1} \bm{\Gamma}_{C_{s(d)}} \bm{B}_C^{-1\dag} \bigg]}{(E^{+}-E_{\alpha})(E^{-}-E_{\beta})},
\end{equation}
with
\begin{equation}
E^{\pm} = E \pm i\epsilon, \epsilon \to 0+,
\end{equation}
and
\begin{equation}
\bm{B}_C = \bm{1}_C-\bm{\Sigma}_C\bm{G}_C^0,
\end{equation}
where $\bm{1}_C$ is the identity matrix of dimension of ($N_C \times N_C$). Note that the dimension of $\bm{\Sigma}_C$ and $\bm{G}_C^0$ is ($N_C \times N_C$) as well.

  \subsection{Incomplete set of CBR eigenstates}\label{sec:methodology:incomplete set of eigenstates}

The major feature of the CBR method is the ability to use a greatly reduced, incomplete set of specially defined eigenstates to represent the true open-system solution. As was shown in \citeauthor{Mamaluy:2003}~\cite{Mamaluy:2003}, it can be accomplished by imposing a special kind boundary condition to the decouple device. The idea is to be able to rewrite  the Green's function matrix of the open system $G_D(E)$, Equation~\ref{eq:Grd}, as
\begin{align}
G_D(E) &= \bigg( \bm{E}^+ - \bm{H}_D^0 - \bm{\Sigma}_s      - \bm{\Sigma}_d \bigg)^{-1} \\\nonumber
       &= \bigg( \bm{E}^+ - \bm{H}_D^N - \bm{\Sigma}_s^N(E) - \bm{\Sigma}_d^N(E) \bigg)^{-1}   
\end{align}
where $\bm{H}_D^N$ is independent of the energy, and $\bm{\Sigma}_s^N(E)$ and $\bm{\Sigma}_d^N(E)$ tend toward zero for values of $E$ that lies not far from the band edge. This enables us to solve the Dyson equation with an incomplete basis.

We start assuming that all nonzero coupling matrix elements of $\bm{W}_{s(d) D}$ is equal to a real constant value, $W_{s(d)}$. This leads to the following expression for the self-energy matrix $\bm{\Sigma}_{s(d)}$ within each source(drain) contact, $s(d)$, see e.g. \citeauthor{Datta:1997}~\cite{Datta:1997}
\begin{equation}\label{eq:self-energy}
\bm{\Sigma}_{s(d)} = - W_{s(d)} \sum_{m}^{M_T} \bm{\chi}_{s(d)}^{m} \exp{ (ik_{s(d)}^{m} a_{s(d)}) } (\bm{\chi}_{s(d)}^{m})^{\dag}
\end{equation}
where $a_{s(d)}=a$ is the constant parameter of the lattice, $M_T$ is the total number of propagating modes, and $\bm{\chi}_{s(d)}^{m}$ are the modes of the Schr\"{o}dinger equation $\bm{H_{s(d)}^{0}}\bm{\chi}_{s(d)}^{m}=\epsilon_{s(d)}^{m}\bm{\chi}_{s(d)}^{m}$. The wave vectors $k_{s(d)}^m$ are functions of energy E as
\begin{equation}
E = \epsilon_{s(d)}^m + 2 W_{s(d)}(1-\cos{(k_{s(d)}^{m}a)})
\end{equation}
The exponential term in Equation~\ref{eq:self-energy} can be approximated in power of $k_{s(d)}^m$ as
\begin{align}
-W_{s(d)}\exp{(ik_{s(d)}^m(E)a)} &=-W_{s(d)}-iW_{s(d)}ak_{s(d)}^m(E)+... \\\nonumber
                                 &=-W_{s(d)}+F_{s(d)}(k_{s(d)}^m(E)),
\end{align}
which leads to
\begin{align}
\bm{\Sigma}_{s(d)}(E) = & -W_{s(d)} \delta_{ij} + \\\nonumber
                        & \sum_m \langle i | \bm{\chi}_{s(d)}^m \rangle F(k_{s(d)}^m(E)) \langle \bm{\chi}_{s(d)}^m|j\rangle ~ i,j \in C_{s(d)}.
\end{align}
Thus, the self-energy matrix in each contact, $s$ and $d$, consists of a first term, which is an independent-energy diagonal matrix, and a second term, that vanishes not too far from the low energy band edge. The total self-energy matrix can be then rewritten as
\begin{equation}
\bm{\Sigma}_{s}(E) + \bm{\Sigma}_{d}(E) = -\bm{K} + \bm{\Sigma}^N(E),
\end{equation}
where
\begin{equation}
\bm{K} =
    \begin{cases}
    	W_{s(d)}\delta_{ij},      \quad &i,j \in    C_{s(d)} \\
    	0,                        \quad &i,j \notin C_{s(d)},\\
    \end{cases}  
\end{equation}
and the retarded Green's function matrix as 
\begin{align}
\bm{G}_D(E) &= [\bm{E}^+-\bm{H}_D^0 - \bm{\Sigma}_s(E) - \bm{\Sigma}_d(E) ]^{-1} \\\nonumber
            &= [\bm{E}^+-\bm{H}_D^0 + \bm{K} - \bm{\Sigma}^N(E)]^{-1}            \\\nonumber
            &= [\bm{E}^+-\bm{H}_D^N - \bm{\Sigma}^N(E)]^{-1}.
\end{align}
The self-energy $\bm{\Sigma}^N(E)$ is small for E close to the band edge, and the Hamiltonian matrix of the device is defined by $\bm{H}_D^N = \bm{H}_D^0 - \bm{K}$. It can be shown that this Hamiltonian matrix corresponds to generalized Neumann boundary conditions for its eigenfunctions \cite{Mamaluy:2005}. Thus, the eigenfunctions of $\bm{H}_D^N$ are approximate solutions of the open-boundary problem in the low-energy limit. As a consequence it suffices to include an incomplete set in the spectral representation of $\bm{H}_D^N$ in the calculation of the open-system quantities of interest within some limited energy range. 

  \subsection{Device Effective-Mass Hamiltonian}\label{sec:methodology:device Effective-Mass Hamiltonian}

The effective mass Schr\"{o}dinger equation for $\Gamma$-valley electrons in Si is given by
\begin{equation}
\bm{H}_D^{0}(\bm{r}_i) \bm{\psi}_\alpha(\bm{r}_i) = E_\alpha \bm{\psi}_\alpha(\bm{r}_i),      
\end{equation}
where $\bm{H}_D^{0}$ is the effective-mass Hamiltonian operator
\begin{align}\label{eq:effective_mass_hamiltonian}
\bm{H}_D^{0}(\bm{r}_i) =& 
-\frac{\hbar^2}{2m_{t}}\frac{\partial^2 }{\partial x^2} 
-\frac{\hbar^2}{2m_{t}}\frac{\partial^2 }{\partial y^2} 
-\frac{\hbar^2}{2m_{l}}\frac{\partial^2 }{\partial z^2} \\\nonumber
&+ \bm{\phi}^H(\bm{r}_i) + \bm{\phi}^{XC}(\bm{r}_i),      
\end{align}
and $\bm{\phi}^H(\bm{r}_i)$ is the Hartree potential and $\bm{\phi}^{XC}(\bm{r}_i)$ is the exchange-correlation potential.

We discretize the 3D domain in $N_x$ equidistant grid-points along the x-axis, in $N_y$ equidistant grid-points along the y-axis and in $N_z$ equidistant grid-points along the z-axis. For all directions, the separation between grid-points is chosen to be ${\Delta}x={\Delta}y={\Delta}z=0.2$~nm. A grid-point is defined by the following triple indices ($i$, $j$, $k$), with $i=1,...,N_x$,  $j=1,...,N_y$ and $k=1,...,N_z$, and the global index $n=k+N_{z}(N_{x}(j-1)+i-1)$. Therefore, the effective mass Hamiltonian matrix for the decoupled device can be expressed with seven non-zero diagonals as follows

\setcounter{MaxMatrixCols}{20}

\begin{align}
\bm{H}_D^{0} = 
\begin{pmatrix}
d_1    & l_1    &         &        & b_1    &        &         &  g_1   &        &        &         \\
l_1    & d_2    & l_2     &        &        & b_2    &         &        & \ddots &        &         \\
       & l_2    & \ddots  &\ddots  &        &        & \ddots  &        &        &  g_n   &         \\
       &        & \ddots  &        &        &        &         & b_n    &        &        & \ddots  \\
b_1    &        &         &        & \ddots & \ddots &         &        &\ddots  &        &         \\
       & b_2    &         &        & \ddots &  d_n   & l_n     &        &        &        &         \\
       &        & \ddots  &        &        &  l_n   & \ddots  & \ddots &        &        &         \\
g_1    &        &         & b_n    &        &        & \ddots  &        &        &        &         \\
       &\ddots  &         &        & \ddots &        &         &        &        &        &         \\
       &        & g_n     &        &        &        &         &        &        &        &         \\
       &        &         & \ddots &        &        &         &        &        &        &        \\
\end{pmatrix},
\end{align}
where
\begin{align}
d_n = & \frac{\hbar^2}{m_l^{(i,j,k)}{\Delta}z^2} + \frac{\hbar^2}{m_t^{(i,j,k)}{\Delta}y^2} +  \frac{\hbar^2}{m_t^{(i,j,k)}{\Delta}x^2},~\text{with}~n=1,...,N_xN_yN_z,
\end{align}
\begin{equation}
l_{n} = \bigg( \frac{1}{m^{(i,j,k)}_l} + \frac{1}{m^{(i,j,k-1)}_l} \bigg) \frac{-\hbar^2}{4{\Delta}z^2},~\text{with}~n=1,...,N_xN_yN_z-1,    
\end{equation}
\begin{equation}
b_{n} = \bigg( \frac{1}{m^{(i,j,k)}_t} + \frac{1}{m^{(i-1,j,k)}_t} \bigg) \frac{-\hbar^2}{4{\Delta}x^2}
,~\text{with}~n=1,...,(N_xN_y-1)N_z,
\end{equation}
and
\begin{equation}
g_{n} = \bigg( \frac{1}{m^{(i,j,k)}_t} + \frac{1}{m^{(i,j-1,k)}_t} \bigg) \frac{-\hbar^2}{4{\Delta}y^2}
,~\text{with}~n=1,...,(N_y-1)N_xN_z,  
\end{equation}
The number of zero diagonals between the diagonals $l$ and $b$ is $(N_z-2)$ and between the diagonals $b$ and $g$ is $(N_x-1)N_z-1$. The exchange-correlation potential $\phi^{XC}(\bm{r}_i)$ that accounts for electron-electron interaction is computed using the terms given in \citeauthor{PerdewZunger:1981}~\cite{PerdewZunger:1981}. 

  \subsection{Charge Self-Consistency}\label{sec:methodology:charge self-consistency}

We solve charge self-consistently the open-system effective mass Schr\"{o}dinger equation and the non-linear Poisson equation \cite{Khan:2007,Gao:2014,Mamaluy:2015}. We employ a single-band effective mass approximation with a valley degeneracy of $d_{val}=6$. For the charge self-consistent solution of the non-linear Poisson equation we use a combination of the predictor-corrector approach and Anderson mixing scheme \cite{Khan:2007,Gao:2014}. 

\begin{figure}
  \centering
  \includegraphics[width=\linewidth]{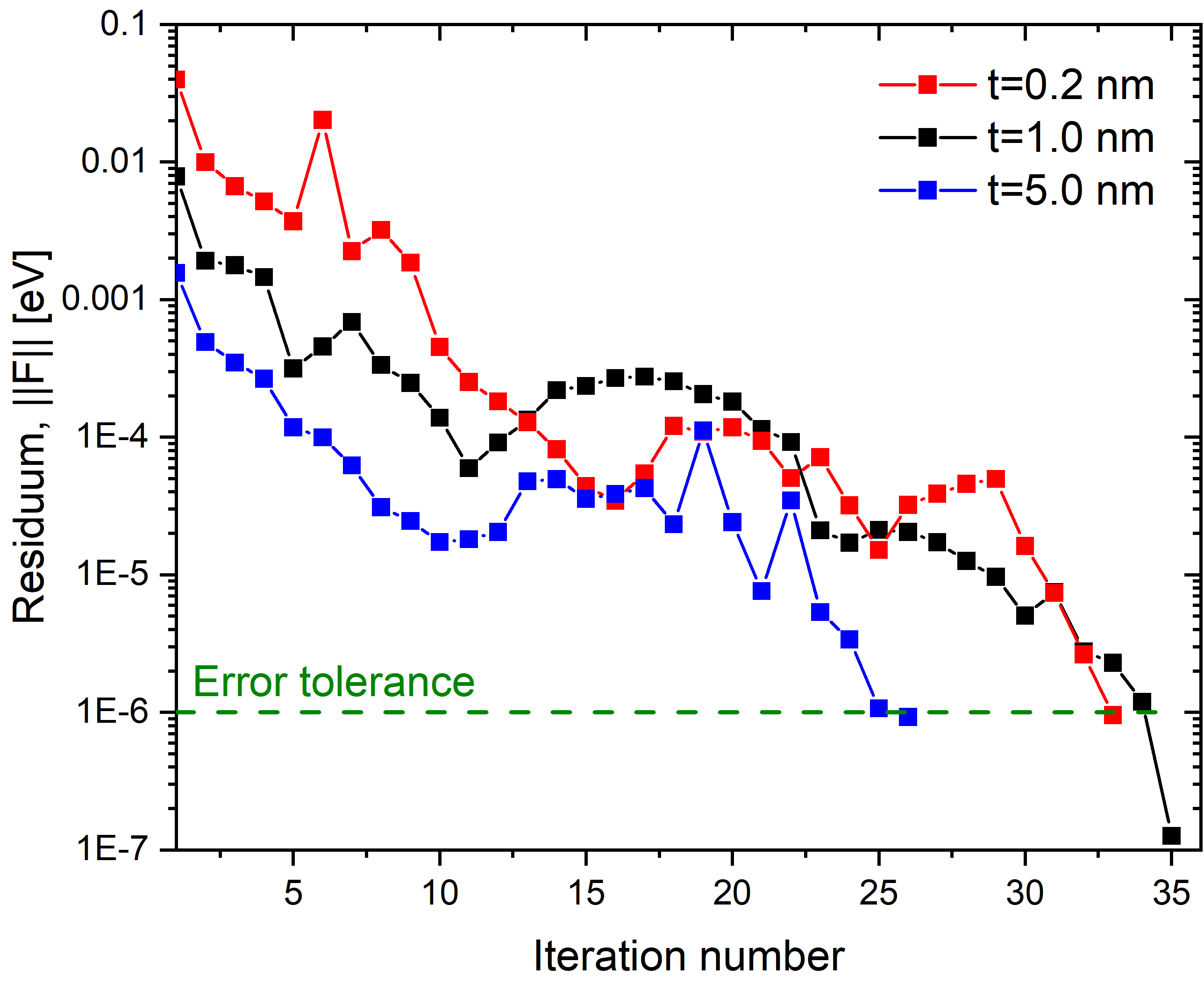}
  \caption{
  \textbf{Convergence evaluation of the method.} The convergence rate of the charge self-consistent open-system Schrodinger-Poisson equations for different $\delta$-layer thicknesses (t) is shown.
  }
  \label{fig:residuum}
\end{figure}

Firstly, the Schr\"{o}dinger equation is solved in a specially defined (see generalized Neumann BC in section~\ref{sec:methodology:incomplete set of eigenstates}) closed-system basis taking into account the Hartree potential $\phi^H(\bm{r}_i)$ and the exchange and correlation potential $\phi^{XC}(\bm{r}_i)$. It is found that out of more than 1,000,000 eigenstates only about 700 ($<0.1\%$) of lowest-energy states is needed (see section \ref{sec:methodology:incomplete set of eigenstates}), which is generally determined by the material properties (e.g. doping level), but not the device size. Then, the LDOS of the open system, $\rho(\bm{r}_i,E)$, and the electron density, $n(\bm{r}_i)$, are computed using the CBR method. The potential and the carrier density are then used to calculate the residuum $F$ of the Posisson equation
\begin{equation}
\big|\big|\bm{F}[\bm{\phi}^H(\bm{r}_i)]\big|\big|=\big|\big|\bm{A}\bm{\phi}^H(\bm{r}_i) - (\bm{n}(\bm{r}_i)-\bm{N}_D(\bm{r}_i)+\bm{N}_A(\bm{r}_i))\big|\big|,
\end{equation}
where $\bm{A}$ is the matrix derived from the discretization of the Poisson equation and $\bm{N}_D$ and $\bm{N}_A$ are the total donor and acceptor doping densities arrays, respectively. If the residuum is larger than a predetermined threshold $\epsilon$, the Hartree potential is updated using the predictor-corrector method, together with the Anderson mixing scheme. Using the updated Hartree potential and the corresponding carrier density, the exchange-correlation is computed again for the next step, and an iteration of Schrodinger-Poisson is repeated until the convergence is reached with $\big|\big|\bm{F}[\bm{\phi}^H(\bm{r}_i)]\big|\big|<\epsilon=10^{-6}$~eV. A typical convergence rate for Si:P $\delta$-layer structures of different thicknesses is shown in Fig.~\ref{fig:residuum}. It illustrates that the proposed charge self-consistent scheme has a robust convergence for different structures, providing the residuum error reduction of 5 orders of magnitude within 40 iterations. In all simulation in this work the standard values of electron effective masses $m_l = 0.98 \times m_e$, $m_t = 0.19 \times m_e$  and the dielectric constant $\epsilon_{Si}=11.7$ of Silicon were employed.

\subsection{From finite-width to infinite-width systems}\label{sec:methodology:from finite-width to infinite-width systems}

In the scenario that the width of a conducting structure along the y-direction is assumed to be infinite (see Fig.~\ref{fig:CBR_model}) with a flat electrostatic potential, we can write the solutions of the Schr\"{o}dinger equation as the product of plane-waves along y-axis and the solutions of the 2D Schr\"{o}dinger equation in the X-Z plane of the device.
\begin{equation}\label{eq:eigenvectors}
\bm{\psi}_{\alpha}(\bm{r}_i)=\frac{\exp{(i k_y y)}}{\sqrt{L_y}} \bm{\psi}_{m}(x,z),
\end{equation}
which is normalized to a length of $L_y$,
and
\begin{equation}\label{eq:eigenvalues}
E_{\alpha}=E_m + \frac{\hbar^2}{2m_t}k_y^2.
\end{equation}
The electron density can be expressed by summing over all (generally - open-system) states $\alpha$ as
\begin{equation}\label{eq:elec_density}
n(\bm{r}_i)=2\sum_\alpha|\bm{\psi}_{\alpha}(\bm{r}_i)|^2f(E_{\alpha}),
\end{equation}
where $k_B=8.617 \times 10^{-5}$~eV~K$^{-1}$ is the Boltzmann constant, $T$ is the temperature of the system and $f(E)$ is the Fermi-Dirac distribution function, which provides the probability that a state is occupied or unoccupied.
The Fermi-Dirac distribution function is given by
\begin{equation}\label{eq:3D_FD}
f(E)= \frac{1}{1+\exp{[E/k_BT]}},
\end{equation}
Using equations~\ref{eq:eigenvectors},~\ref{eq:eigenvalues}~and~\ref{eq:elec_density} we get
\begin{align} \label{eq:FD_sum}
n(\bm{r}_i) &=2\sum_{\alpha} |\bm{\psi}_{\alpha}(x,z)|^2f(E_{\alpha})  \\\nonumber
            &= 2 \sum_m |\bm{\psi}_m(x,z)|^2 \bigg( \frac{1}{L_y} \sum_{k_y} f( E_m + \frac{\hbar^2}{2m_t}k_y^2) \bigg) \\\nonumber
            &=2 \sum_m |\bm{\psi}_m(x,z)|^2 f^{2D}( E_m ),
\end{align}
where we denote the term in parenthesis as the effective 2D Fermi-Dirac distribution function $f^{2D}(E)$. Replacing the summation over $k_y$ with an integral as $\frac{1}{L_y}\sum_{k_y} \rightarrow \int_{-\infty}^{\infty} \frac{dk_y}{2\pi}$, we can get the distribution function in an integral form as follows
\begin{align}
f^{2D}( E ) &= \frac{1}{L_y} \sum_{k_y} f( E + \frac{\hbar^2}{2m_t}k_y^2)  \\\nonumber
            &= \int_{-\infty}^{\infty} \frac{1}{2\pi} f( E + \frac{\hbar^2}{2m_t}k_y^2) dk_y \\\nonumber
            &= \int_{0}^{\infty} \frac{1}{\pi} f( E + \frac{\hbar^2}{2m_t}k_y^2) dk_y   \\\nonumber
            &= \int_{0}^{\infty} \frac{1}{\pi} \frac{1}{ 1+\exp{\frac{E}{k_BT}}\exp{ \big( \frac{\hbar^2k_y^2}{2 m_t k_B T}} \big) } dk_y.
\end{align}
Finally, performing a variable change $l=\frac{\hbar^2 k_y^2}{2m_t k_B T}$ we obtain
\begin{align}\label{eq:effective_2D_FD}
f^{2D}( E ) &=\bigg( \frac{m_t k_B T}{ 2 \pi \hbar^2} \bigg)^{1/2} \frac{1}{\sqrt{\pi}} \int_0^{\infty} \frac{1}{1+\exp{(\frac{E}{k_BT}+l})} \frac{1}{\sqrt{l}}dl \\\nonumber
&=
\bigg( \frac{m_t k_B T}{ 2 \pi \hbar^2} \bigg)^{1/2} f_{-1/2} \bigg( -\frac{E}{k_BT} \bigg),
\end{align}
where $f_{-1/2}(x)$ is the Fermi-Dirac Integral of the order of $-1/2$. It can be expressed, for instance, through the polylogarirthm  special function as $f_{-1/2}(x)=-\textrm{Li}_{1/2}(-\exp(x))$~\cite{Polylog}.

\begin{figure}
  \centering
  \includegraphics[width=\linewidth]{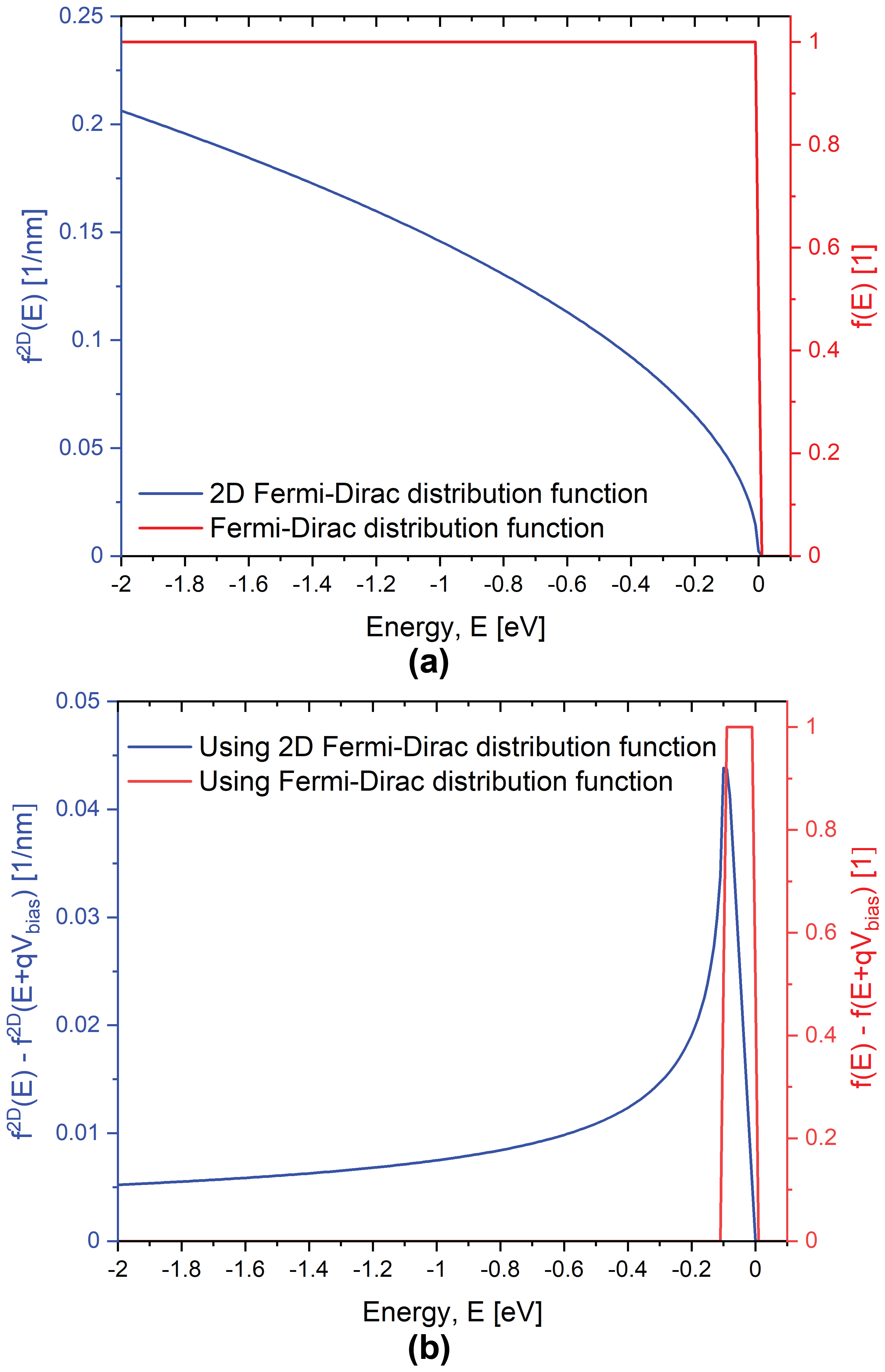}
  \caption{
  \textbf{Fermi-Dirac distribution functions.}
  \textbf{a} Comparison between the Fermi-Dirac distribution function, Eq.~\ref{eq:3D_FD}, and the effective 2D Fermi-Dirac distribution function, Eq.~\ref{eq:effective_2D_FD}, for equilibrium conditions and $T=4.0$~K. \textbf{b} Comparison between the Fermi-Dirac function difference between source and drain, $f_{s}(E)-f_{d}(E)=f(E)-f(E+qV_{bias})$, and the corresponding effective 2D Fermi-Dirac distribution function, $f_{s}^{2D}(E)-f_{d}^{2D}(E)=f^{2D}(E)-f^{2D}(E+qV_{bias})$, for an applied voltage of $V_{bias}=0.1$~V and $T=4.0$~K.
  }
  \label{fig:FD_distribution_function}
\end{figure}

Fig.~\ref{fig:FD_distribution_function}a shows the Fermi-Dirac distribution function, Eq.~\ref{eq:3D_FD}, and the effective 2D Fermi-Dirac distribution function, Eq.~\ref{eq:effective_2D_FD}, in equilibrium conditions and a temperature of $4.0$~K. In both cases the occupied states exists only below the Fermi level, but the occupation rate for the effective 2D distributions increases as $\sim\sqrt{-E}$ for $E\to-\infty$. We next examine the behaviour of the current integral given in the Eq.~\ref{eq:current} at low temperatures. In general, the difference in the distribution functions  $f_s(E)-f_d(E) = f(E)-f(E+qV_{bias})$ determines the energy range of non-zero contributions to the total current. The difference between Fermi-Dirac distributions is non-zero only in a small range of energies in the vicinity of the Fermi level as the result in red line in Fig.~\ref{fig:FD_distribution_function}b indicates. Furthermore, the energy range for non-zero contribution is proportional to the applied voltage $V_{bias}=0.1$~V. On the contrary, the difference in the effective 2D distribution functions is non-zero for all energies below the Fermi level as the plot in blue color in Fig.~\ref{fig:FD_distribution_function}b illustrates. This is a reflection of the asymptotic behaviour of $f^{2D}(E)$ at $E\to-\infty$; for the difference in the distributions one gets $f_{s}^{2D}(E)-f_{d}^{2D}(E) = f^{2D}(E)-f^{2D}(E+qV_{bias})\sim qV_{bias}/\sqrt{-E}$.

We thus conclude that, within the Landauer-Buttiker/NEGF formalism, finite size structures in the transverse directions have a principally different current spectrum from the structures that are infinite along a transverse direction (e.g. when $W\to\infty$ in $\delta$-layer structures shown in Fig.~\ref{fig:ideal_TJ_model}). 

\section*{acknowledgement}\label{sec:acknowledgement}

This work is funded under Laboratory Directed Research and development (LDRD) program, Project No. 227155, at Sandia National Laboratories. Sandia National Laboratories is a multimission laboratory managed and operated by National Technology and Engineering Solutions of Sandia, LLC., a wholly owned subsidiary of Honeywell International, Inc., for the U.S. Department of Energy’s National Nuclear Security Administration under contract DE-NA-0003525. This paper describes objective technical results and analysis. Any subjective views or opinions that might be expressed in the paper do not necessarily represent the views of the U.S. Department of Energy or the United States Government.

This article has been authored by an employee of National Technology $\&$ Engineering Solutions of Sandia, LLC under Contract No. DE-NA0003525 with the U.S. Department of Energy (DOE). The employee owns all right, title and interest in and to the article and is solely responsible for its contents. The United States Government retains and the publisher, by accepting the article for publication, acknowledges that the United States Government retains a non-exclusive, paid-up, irrevocable, world-wide license to publish or reproduce the published form of this article or allow others to do so, for United States Government purposes. The DOE will provide public access to these results of federally sponsored research in accordance with the DOE Public Access Plan https://www.energy.gov/downloads/doe-public-access-plan.

This version of the article has been accepted for publication, after peer review (when applicable) but is not the Version of Record and does not reflect post-acceptance improvements, or any corrections. The Version of Record is available online at: http://dx.doi.org/10.1038/s41598-022-20105-x

\section*{Author contributions}
J.P.M. and D.M. performed equally the central calculations, analysis and discussion presented in this work. The manuscript was written by all the authors.

\section*{Data availability}
The data that support the plots within this paper are available from the corresponding authors upon reasonable request.

\section*{Competing interests}
The authors declare no competing interests.

\bibliographystyle{apsrev4-2}
\bibliography{references}

\end{document}